\documentclass{article}

\usepackage{arxiv}

\usepackage[utf8]{inputenc} 
\usepackage[T1]{fontenc}    
\usepackage{hyperref}       
\usepackage{url}            
\usepackage{booktabs}       
\usepackage{amsfonts}       
\usepackage{nicefrac}       
\usepackage{microtype}      
\usepackage{lipsum}

\usepackage{subcaption}
\usepackage{graphicx}
\usepackage{multirow}
\usepackage{amsmath}
\usepackage{bm}
\usepackage{pifont}
\newcommand{\cmark}{\text{\ding{51}}}
\newcommand{\xmark}{\text{\ding{55}}}
\usepackage{authblk}

\newcommand*{\affaddr}[1]{#1} 
\newcommand*{\affmark}[1][*]{\textsuperscript{#1}}
\newcommand*{\email}[1]{\texttt{#1}}

\title{The 2ST-UNet for Pneumothorax Segmentation in Chest X-Rays using ResNet34 as a Backbone for U-Net}

\author{Ayat Abedalla\affmark[1],
Malak Abdullah\affmark[1],
Mahmoud Al-Ayyoub\affmark[1], and
Elhadj Benkhelifa\affmark[2] \\ 
\affaddr{\affmark[1]Jordan University of Science and Technology, Irbid, Jordan}\\
\affaddr{\affmark[2]Staffordshire University, Stoke-on-Trent, UK}\\
\email{ayatabedalla@gmail.com,
mabdullah@just.edu.jo,
maalshbool@just.edu.jo,
E.Benkhelifa@staffs.ac.uk}}

\begin{document}
\maketitle

\begin{abstract}

Pneumothorax, also called a collapsed lung, refers to the presence of the air in the pleural space between the lung and chest wall. It can be small (no need for treatment), or large and causes death if it is not identified and treated on time. It is easily seen and identified by experts using a chest X-ray. Although this method is mostly error-free, it is time-consuming and needs expert radiologists. Recently, Computer Vision has been providing great assistance in detecting and segmenting pneumothorax. In this paper, we propose a 2-Stage Training system (2ST-UNet) to segment images with pneumothorax. This system is built based on U-Net with Residual Networks (ResNet-34) backbone that is pre-trained on the ImageNet dataset. We start with training the network at a lower resolution before we load the trained model weights to retrain the network with a higher resolution. Moreover, we utilize different techniques including Stochastic Weight Averaging (SWA), data augmentation, and Test-Time Augmentation (TTA). We use the chest X-ray dataset that is provided by the 2019 SIIM-ACR Pneumothorax Segmentation Challenge, which contains 12,047 training images and 3,205 testing images. Our experiments show that 2-Stage Training leads to better and faster network convergence. Our method achieves 0.8356 mean Dice Similarity Coefficient (DSC) placing it among the top 9\% of models with a rank of 124 out of 1,475.
\end{abstract}

\keywords{Pneumothorax Segmentation \and 2-Stage Training \and U-Net \and ResNet-34 \and Chest X-ray \and Transfer Learning \and Data Augmentation \and Test-Time Augmentation}

\section{Introduction}
\begin{figure}
\centering
\includegraphics[width=5cm]{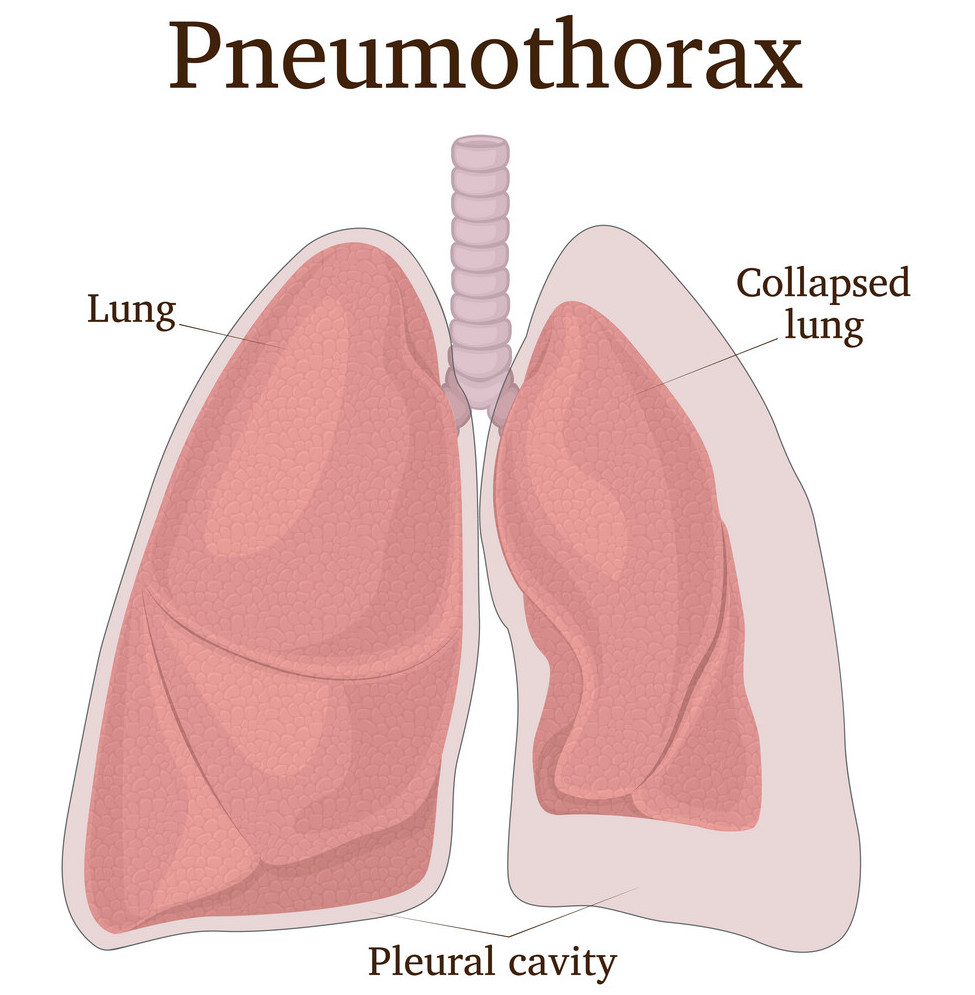}
\caption{Illustration of Pneumothorax.\protect\footnotemark}
\label{fig:pnue}
\end{figure}

Pneumothorax (Collapsed Lung) is the presence of air in the pleural cavity between the lungs and the chest wall. The pressure of this air causes the lung to collapse on itself~\cite{intro_pneumothorax2}, see Figure~\ref{fig:pnue} for more illustration. The sudden chest pain and shortness of breath are considered the main symptoms of pneumothorax. It can be caused by a variety of reasons, such as lung diseases or defects, and in some cases, it can occur from an accident or injury in the chest area, while, in other cases, it may occur without any apparent reason. Pneumothorax can be small (no need for treatment), or large and causes death if it is not identified and treated immediately (subsequent dyspnea and alveoli explosion due to the presence of air)~\cite{intro_pleural}. Detecting pneumothorax is complicated due to the variety of its symptoms and causes. It is usually detected by a radiologist using chest X-rays. However, it can be difficult to diagnose especially when its locations are atypical or when the patient has heart or lung diseases~\cite{intro_pneumothorax1}.

Machine Learning and Computer Vision techniques have led to remarkable advances in medical image analysis. Recently, a branch of machine learning, known as deep learning, has been used to process massive amounts of data to automatically learn their features without the need for manual feature extraction/engineering~\cite{al2019detailed}. One of the modern deep learning models is known as Convolutional Neural Networks (CNN)~\cite{cnn}, which achieved great success in general and biomedical image analysis~\cite{proc_rle}.

\footnotetext{\url{https://www.vectorstock.com/royalty-free-vector/pneumothorax-vector-23029312}}

Image segmentation techniques, which aim to partition a digital image into parts or image objects, are extremely useful for several tasks. For medical image processing, there is a need to localize and segment objects or regions, such as brain tumors~\cite{tumor1,tumor2}, brain tissues~\cite{tissue}, and abdominal aortic aneurysms~\cite{aneurysms}. Recently, Fully Convolutional Networks (FCN)~\cite{fcn} have shown great success in medical image segmentation. This success has been mainly associated with efficient architectures, such as SegNet\cite{segnet} and U-Net~\cite{unet}. 

In this study, we propose a 2-Stage Training method to segment pneumothorax in chest X-rays. The dataset we use was provided by the SIIM-ACR Pneumothorax Segmentation Challenge.\footnote{\url{https://www.kaggle.com/c/siim-acr-pneumothorax-segmentation}} The proposed segmentation model has a U-Net~\cite{unet} architecture with 34-layer Residual Network (ResNet-34)~\cite{resnet} pre-trained on the ImageNet~\cite{imagenet} as a backbone. We have resized the original $1024 \times 1024$ X-ray images to $256 \times 256$ and $512 \times 512$ image sizes. In stage-1, the model is trained on the lower resolution of $256 \times 256$. Then, the weights of the model are loaded in stage-2 to train the model on the higher resolution of $512 \times 512$. As a result, the model achieves 0.8356 mean Dice Similarity Coefficient (DSC)~\cite{dice}.

This paper is organized as follows. Section~\ref{sec:related} presents the related work on pneumothorax research and the top teams for the SIIM-ACR Pneumothorax Segmentation Challenge. Section~\ref{sec:data} describes the dataset that is used in this paper. Section~\ref{sec:method} provides an overview of our methodology and the architecture of our approach. Section~\ref{sec:exp} presents experimental settings and metrics for evaluations. The results and discussion are given in Section~\ref{sec:res}. Finally, the conclusion and future work are provided in Section~\ref{sec:conc}.

\section{Related Work}
\label{sec:related}

With the advancement of deep learning techniques in medical image processing, the tasks of classification, detection, and segmentation of medical images have attracted the attention of many researchers. In the classification task, the model takes the entire image and maps it into discrete class labels. The classification task of images is one of the core problems in which deep learning has made a great contribution, especially in the field of medical image analysis. In~\cite{diabetic_class}, the researchers deployed deep Convolutional Neural Network (DCNN) with utilizing a dropout technique for diabetic retinopathy classification from a fundus image. They evaluated on Kaggle, DRIVE, and STARE public datasets with an accuracy of 94–96\%. In~\cite{seg_based}, the researchers proposed an automatic knowledge extraction from chest X-ray images. First, they compared five classifiers of traditional methods and five CNN-based methods. The traditional methods achieve 0.52 of the area under the receiver operating characteristic (AUROC) and CNN achieves 0.96 scores.

The detection task allows us to identify and locate the objects or regions of interest in an image by creating a bounding box around it. Deep learning has great potentials in the detection fields. This helps in the early discovery of some diseases, which is one of the important tasks in the diagnosis. Researchers in~\cite{liu2017detecting} built a framework based on CNN architecture for detecting and localizing tumors in breast cancer. They used the Inception V3~\cite{inception} model and evaluated on gigapixel images from the Camelyon16 dataset.\footnote{\url{https://camelyon16.grand-challenge.org/Home/}} Their framework could reduce false-negative rates in metastasis detection. In~\cite{clu_detect}, the researchers developed a framework that aimed to improve domain adaptation capability without specific domain adaptation training, called Clustering Convolutional Neural Networks (CLU-CNNs). This framework had a domain adaptation mechanism called Agglomerative Nesting Clustering Filtering (ANCF) and BN-IN Net that is embedded in FCN to extract feature maps.

\begin{figure}
\centering
\includegraphics[width=9cm]{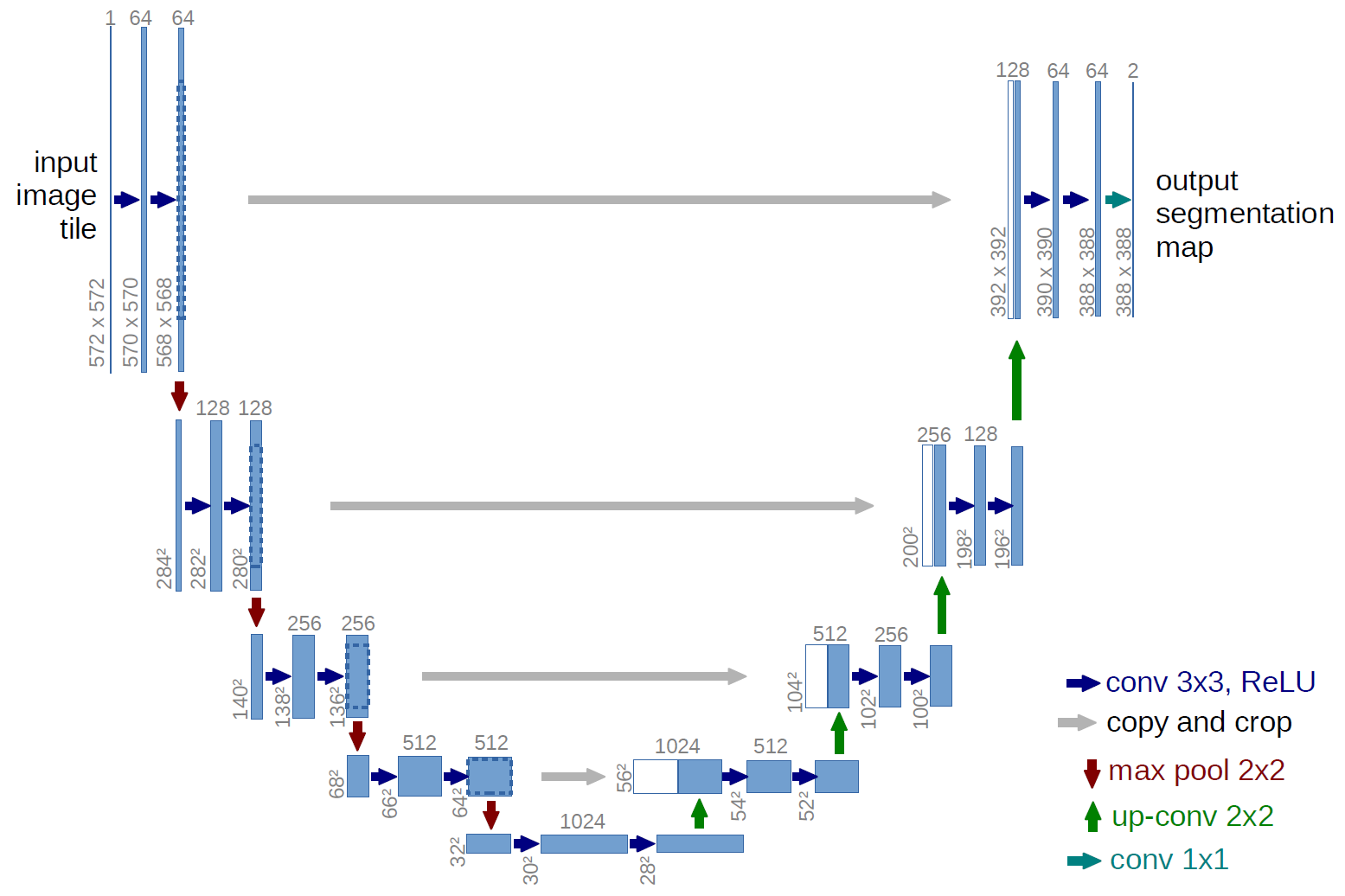}
\caption{U-Net Architecture~\cite{unet}.}
\label{fig:unet}
\end{figure}

Knowing that the segmentation technique aims to partition an image into various meaningful parts (segments) that have similar characteristics, this has helped to understand the objects in the image. Deep learning has shown a great leap in model performance for segmentation tasks. Great success can be seen in the semantic segmentation task, which aims to label and classify each pixel in the image for its own category. This led to special enhancements in the analysis, diagnosis, and treatment in the medical imaging field, such as detecting tissue or organs~\cite{med_seg2018}. A group of researchers~\cite{cnn_tumorbrain} proposed a CNN model with utilizing small kernels for brain tumors segmentation in Magnetic Resonance Imaging (MRI). Their approach achieved first place on a 2013 Brain Tumor Segmentation (BRATS) Challenge, and second place on the 2015 BRATS Challenge. Moreover, a research paper~\cite{unet} introduced an end-to-end semantic segmentation network for biomedical images, called U-Net (see Figure~\ref{fig:unet} for more illustration). This network achieved promising results for neuronal structures segmentation and cell segmentation. Due to the advanced performance of the U-Net, it has become a popular network for medical image segmentation tasks, such as lung segmentation from CT scans~\cite{unet_lung}, retinal layers segmentation from eye optical coherence tomography (OCT) scans~\cite{unet_retinal_eye}, and brain tumor segmentation from MRI images~\cite{unet_brain}.

Pneumothorax, as a medical condition, has attracted the attention of many researchers for detecting and segmenting its medical images. Pneumothorax is a lung pathology that can be life-threatening and considered a medical and surgical emergency, so it is important to have algorithms to automatically detect the problem to provide the proper treatment. In~\cite{rel1}, the authors evaluated the performance of three deep learning techniques: CNN, FCN, and Multiple-Instance Learning (MIL)~\cite{rle2}, to detect and localize pneumothorax in chest X-ray images. They found that CNN had the best performance in terms of Area Under The Curve (AUC), while the other models detect the location of pneumothorax more accurately. In~\cite{rle3}, the researchers used DCNN to identify moderate and large pneumothorax in chest X-rays images. 

On August 21, 2019, a 2-round competition for identifying pneumothorax disease in chest x-rays was released, called SIIM-ACR Pneumothorax Segmentation Challenge. In round-1, each participating team developed a model to segment pneumothorax based on the provided training set. The number of participants in round-1 was 1,475 teams. In round-2, each team had to use the model they developed in round-1 and retrain it on an updated training data set. Then, they had to use the retrained models for inferring and predicting an unseen test data set. The models of the top ten winning teams\footnote{The top 10 winning teams models of the Pneumothorax Challenge are available on \url{https://siim.org/page/pneumothorax_challenge}} rely mostly on deep CNN. The first-place team, Team [dsmlkz] sneddy, used U-Net model with ResNet-34, ResNet-50, and 50-layer ResNext~\cite{resnext} with the Squeeze-and-Excitation (SE) module (SE-ResNext-50)~\cite{squeeze} . The models were trained several times. First, these models (except for ResNet-50) were trained on size $512 \times 512$ then uptrained on size $1024 \times 1024$ with a frozen encoder in early epochs. The best model from the previous step was uptrained until convergence with a 1e-5 learning rate and 0.6 sample rate. Then, they repeated the previous step with a 0.4 sample rate. Finally, the models were uptrained with a small learning rate of 1e-6 and a sample rate of 0.5. They also used aggressive augmentation for training and applied horizontal flip Test-Time Augmentation (TTA)~\cite{tta} and triplet threshold for post-processing. The second-place team, Team X5, used classification and segmentation models in their approach. For classification, they used stacking ensemble of multi-task U-Net model with SE-ResNeXt-50, SE-ResNeXt-101 and EfficientNet-b3~\cite{efficientnet} backbones. For segmentation, they used an averaged ensemble of U-Net model with SE-ResNeXt-50, SE-ResNeXt-101, EfficientNet-b3, EfficientNet-b5 backbones and DeepLabv3~\cite{deeplabv3} with ResNeXt-50 backbone. 
In this work, we have used the same dataset of the SIMM-ACR competition to segment and detect Pneumothorax.

\section{Dataset}
\label{sec:data}

\begin{figure}
\centering
\includegraphics[width=8cm]{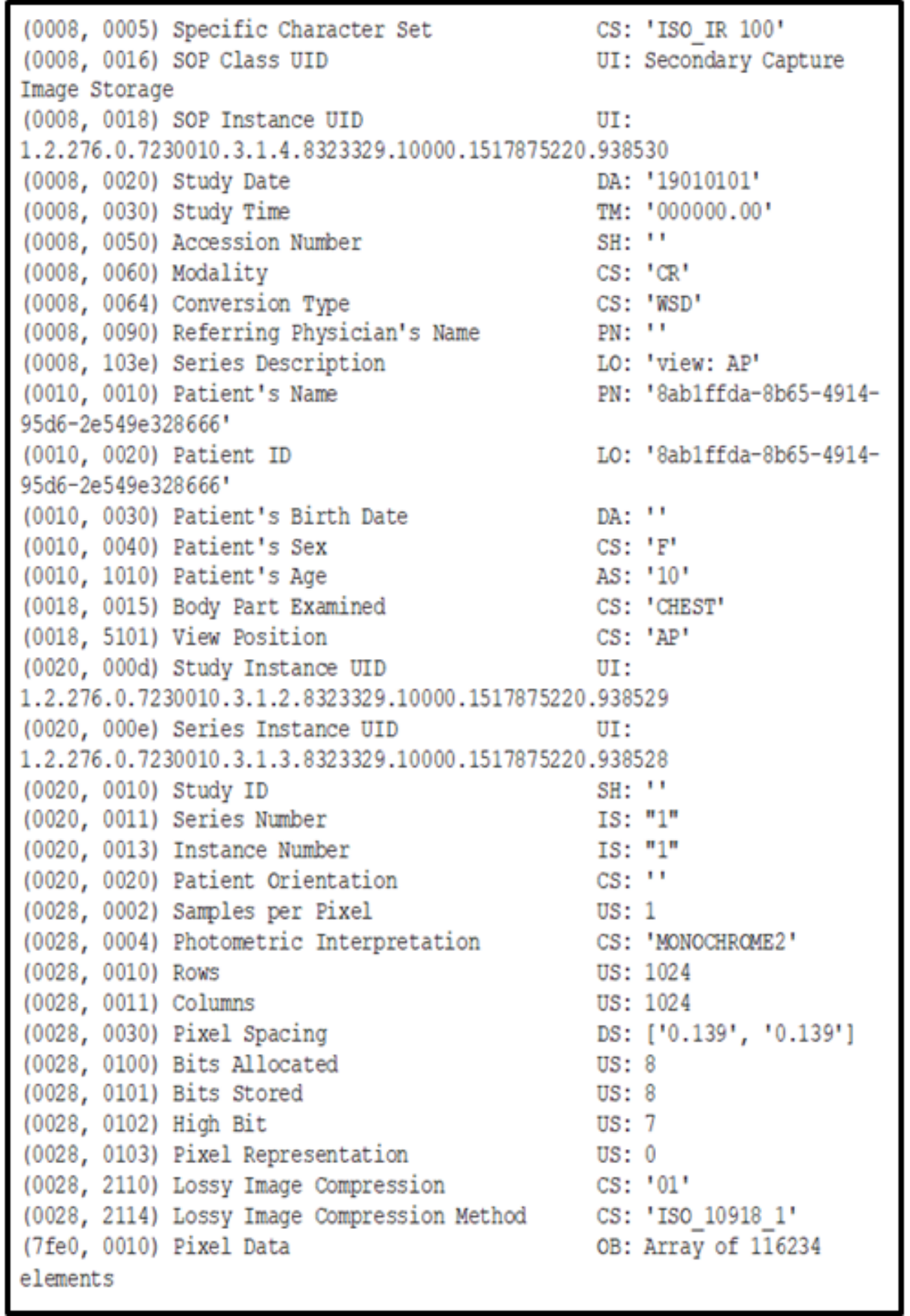}
\caption{Data elements in DICOM file.}
\label{fig:dicom}
\end{figure}

\begin{figure}
\centering
\includegraphics[width=0.9\textwidth]{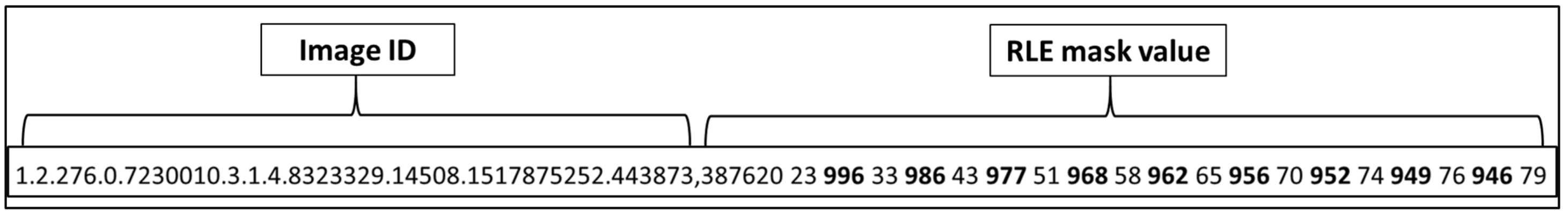}
\caption{Example of RLE.}
\label{fig:rle}
\end{figure}

In this study, we have used the chest X-ray dataset from the SIIM-ACR Pneumothorax Segmentation Challenge. The datasets are obtained by Cloud Healthcare API,\footnote{\url{https://cloud.google.com/healthcare/}} the images of this dataset were released in the Digital Imaging and Communications in Medicine (DICOM)\footnote{\url{https://www.dicomstandard.org/current/}} format, while the annotations are stored as Fast Healthcare Interoperability Resources (FHIR).\footnote{\url{https://www.hl7.org/fhir/index.html}} The DICOM format can store medical images (pixel data) along with patient information (metadata) in one file. Figure~\ref{fig:dicom} provides an example of data elements in the DICOM file. Table~\ref{tab:metadatstat} provides an analysis of the gender and view position metadata for round-1 and round-2 dataset. The gender of the patients is either male or female and the view position is either Anterior-Posterior (AP) frontal X-ray or Posterior-Anterior (PA) frontal X-ray. From the FHIR datastore, we access the annotations mask for the training dataset. The annotation in the form of image IDs and Run-Length Encoding (RLE)\footnote{\url{https://en.wikipedia.org/wiki/Run-length_encoding}} masks. RLE is a form of lossless data compression in which runs of data are stored as a single data value and count. The relative form of RLE is used for images with pneumothorax as a mask value where pixel locations are measured from the previous end of the run as shown in Figure~\ref{fig:rle}. On the other hand, images without pneumothorax have a mask value of -1. All images are in the size of $1024 \times 1024$ pixels. 

\begin{table}
\caption{An analysis of gender and view position metadata for the round-1 and round-2 dataset.}
\label{tab:metadatstat}
\centering
\begin{tabular}{cccccc}
\toprule
\textbf{Dataset} & \textbf{Attribute} & \textbf{Value} & \textbf{Pneumothorax} & \textbf{Healthy} & \textbf{Total}\\
\midrule

\multirow{4}{*}{Round-1 Training} & 
\multirow{2}{*}{Gender} & Female & 1053 (22.0\%) & 3742 (78.0\%) & 4795 (44.9\%) \\
& & Male & 1326 (22.6\%) & 4554 (77.4\%) & 5880 (55.1\%)\\
\cmidrule{2-6}

& \multirow{2}{*}{View Position} & AP & 858 (20.5\%) & 3323 (79.5\%) & 4181 (39.2\%)\\
& & PA & 1521 (23.4\%) & 4973 (76.6\%) & 6494 (60.8\%)\\
\midrule

\multirow{4}{*}{Round-1 Test} & 
\multirow{2}{*}{Gender} & Female & - & - & 626 (45.6\%) \\
& & Male & - & - & 746 (54.4\%) \\
\cmidrule{2-6}

& \multirow{2}{*}{View Position} & AP & - & - & 592 (43.1\%) \\
& & PA & - & - & 780 (56.9\%) \\
\midrule

\multirow{4}{*}{Round-2 Training} & 
\multirow{2}{*}{Gender} & Female & 1180 (21.8\%) & 4241 (78.2\%) & 5421 (45.0\%) \\
& & Male & 1489 (22.5\%) & 5137 (77.5\%) & 6626 (55.0\%)\\
\cmidrule{2-6}

& \multirow{2}{*}{View Position} & AP & 972	(23.3\%) & 3801 (76.7\%) & 4773 (39.6\%)\\
& & PA & 1697 (20.4\%) & 5577 (79.6\%) & 7274 (60.4\%)\\
\midrule

\multirow{4}{*}{Round-2 Test} & 
\multirow{2}{*}{Gender} & Female & - & - & 1452 (45.3\%) \\
& & Male & - & - & 1753 (54.7\%) \\
\cmidrule{2-6}

& \multirow{2}{*}{View Position} & AP & - & - & 1244 (38.8\%) \\
& & PA & - & - & 1961 (61.2\%) \\

\bottomrule\\
\end{tabular}
\end{table}

As mentioned earlier, this challenge had two rounds. The round-1 dataset consists of 10,675 training chest radiographs with 11,582 masks and 1,372 test chest radiographs. The round-2 dataset consists of 12,047 training chest radiographs with 12,954 masks and 3,205 test chest radiographs. The round-2 training set is a combination of the round-1 training set and the round-1 test set with updated labels. In the training set for both rounds, there are more masks than the images. This indicates that some training images have multiple annotations, i.e., multiple regions of interest. The true masks for the round-2 test dataset were not provided, which means we need to submit our prediction results online to evaluate our model.

\begin{figure}
\centering
\includegraphics[width=0.9\textwidth]{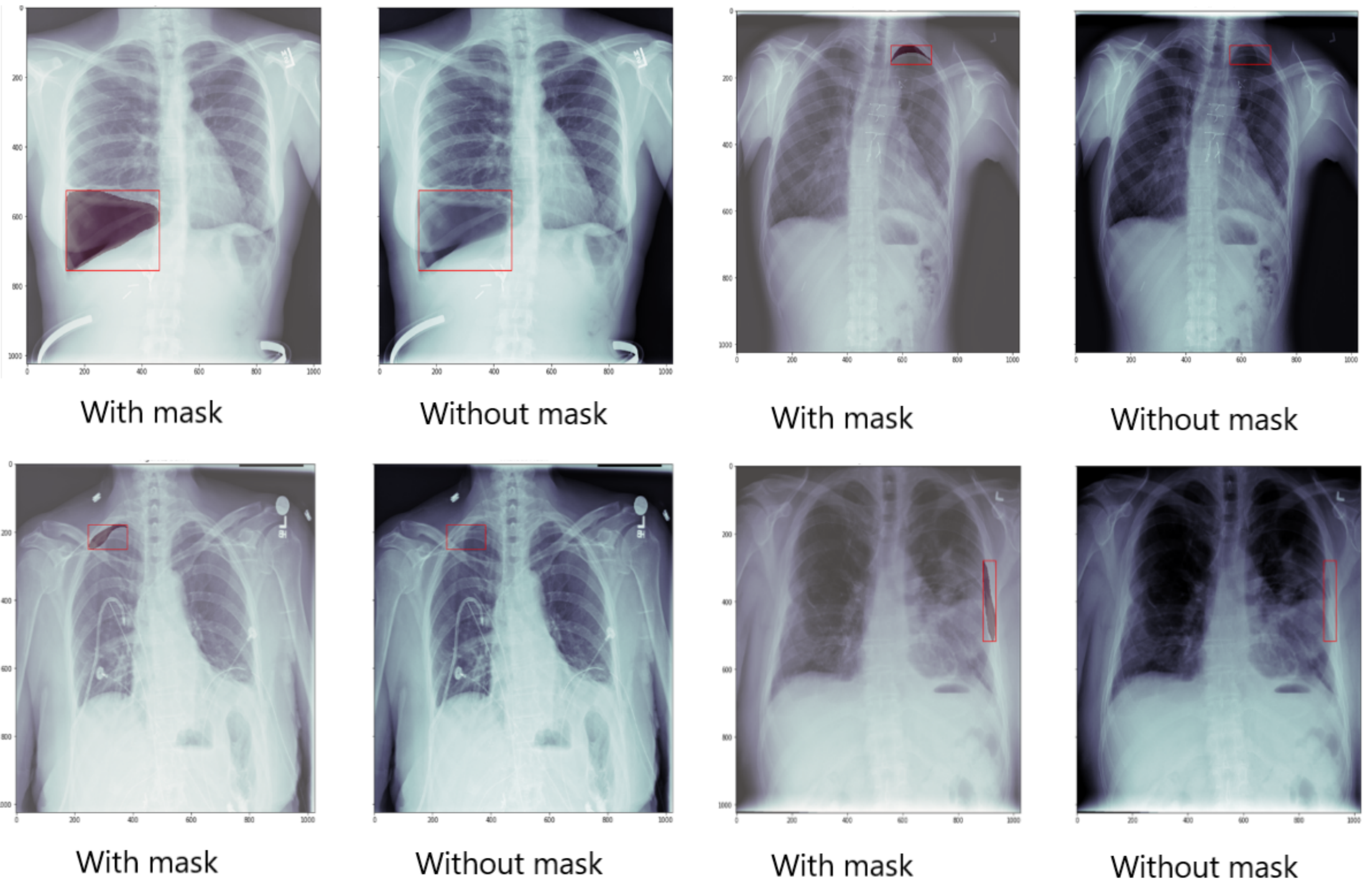}
\caption{Samples of chest X-ray image containing pneumothorax.}
\label{fig:img_mask}
\end{figure}

The round-1 training set contains 2,379 positive cases of finding pneumothorax, while the round-2 training set has 2,669 positive cases of finding pneumothorax. Table~\ref{tab:attributes} shows more details for the dataset. RLE masks for the positive cases in the round-1 training set are distributed as follows: 1,755 cases have a single annotation (73.77\%) and 624 cases have multiple annotations (26.23\%). Whereas in the round- training set, 2,045 cases contain a single annotation (76.62\%) and 624 cases contain multiple annotations (23.38\%). Figure~\ref{fig:img_mask} shows some X-ray images that contain pneumothorax.

\begin{table}
\caption{The overview of dataset statistics for the pneumothorax challenge in Round-1 and Round-2.}
\label{tab:attributes}
\centering
\begin{tabular}{ccc}
\toprule
\textbf{Attribute} & \textbf{Round-1} & \textbf{Round-2} \\
\midrule

Number of training samples & 10,675 & 12,047 \\
Number of test samples & 1,372 & 3,205	 \\
Number of positive cases in training set & 2,379 & 2,669 \\
Number of negative cases in training set & 8,296 & 9,378 \\
Number of cases has single mask in training set & 1,755 & 2,045 \\
Number of cases has multiple masks in training set & 624 & 624 \\

\bottomrule\\
\end{tabular}
\end{table}

\section{Proposed Approach}
\label{sec:method}
\begin{figure}
\centering
\includegraphics[width=0.9\textwidth]{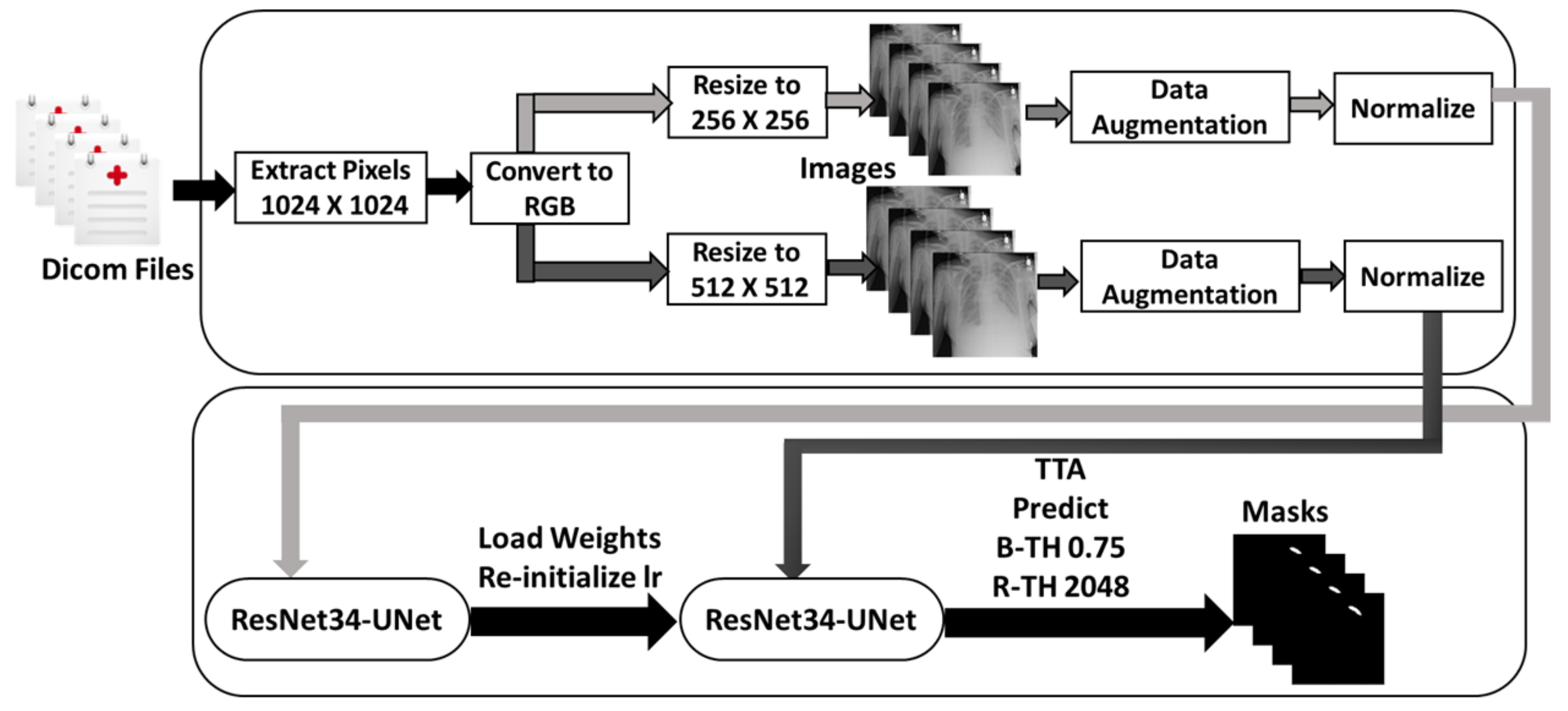}
\caption{Our methodology for pneumothorax segmentation.}
\label{fig:method}
\end{figure}

In this section, we provide details about the image pre-processing and augmentation techniques that have been used in our approach. Then, we present the segmentation network, training, and testing details. The detailed methodology of our approach is illustrated in Figure~\ref{fig:method}.

\subsection{Data Pre-processing and Augmentation}

First, we extract the pixel array in a resolution of $1024 \times 1024$ pixels from the DICOM files and convert the grayscale images (that have only one channel) to RGB colored images (with three channels). Then, we resize these images to $256 \times 256$ pixels and $512 \times 512$ pixels for the 2-Stage Training steps. Also, we extract the $1024 \times 1024$ segmentation maps (masks) from RLE and resize them to $256 \times 256$ and $512 \times 512$. The mask values are 0 for pixels without pneumothorax and 255 for pixels with pneumothorax. 

Data augmentation is a useful way to reduce the generalization error (overfitting) of models by increasing the amount of training data and adding a variety of distortions and noise to the training data. This aims to strengthen and enhance the robustness of the model~\cite{ebrahim2018performance}. We have applied four groups of data augmentation techniques on images and segmentation maps. These techniques include horizontal flip, one of random contrast, random gamma, and random brightness, one of elastic transform, grid distortion, and optical distortion, and random sized crop. Finally, we have normalized the pixel values of images so the range of pixel intensity values are between 0 and 1 and normalized the mask values to 0 or 1 instead of 0 or 255.

\subsection{Model Architecture}

In our approach, we use U-Net~\cite{unet} as a segmentation model, which is an extension of FCN. The backbone used is ResNet-34~\cite{resnet} pre-trained on the ImageNet~\cite{imagenet} dataset. This is an example of transfer learning, which is an effective technique where a model that is trained for one problem can be reused as initialization for another model that is to be trained on another similar problem~\cite{ebrahim2019will}. The network architecture is based on a 2-Stage Training technique. First, we have trained our network on lower resolution images ($256 \times 256$). Then, the model's weights are loaded and the learning rate is initialized to train the model on higher resolution ($512 \times 512$).

\begin{figure}[t!]
 \centering
 \begin{subfigure}[t]{0.49\textwidth}
 \centering
 \includegraphics[height=3.5in]{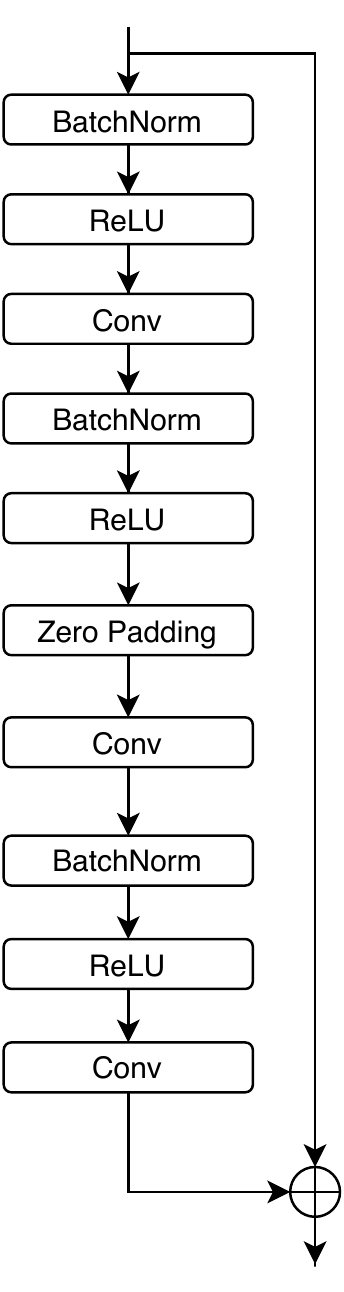}
 \caption{ResNet-E1 is unit 1 of the encoder part.}
 \label{fig:ch4_block_details_a}
 \end{subfigure}%
 ~ 
 \begin{subfigure}[t]{0.49\textwidth}
 \centering
 \includegraphics[height=3.5in]{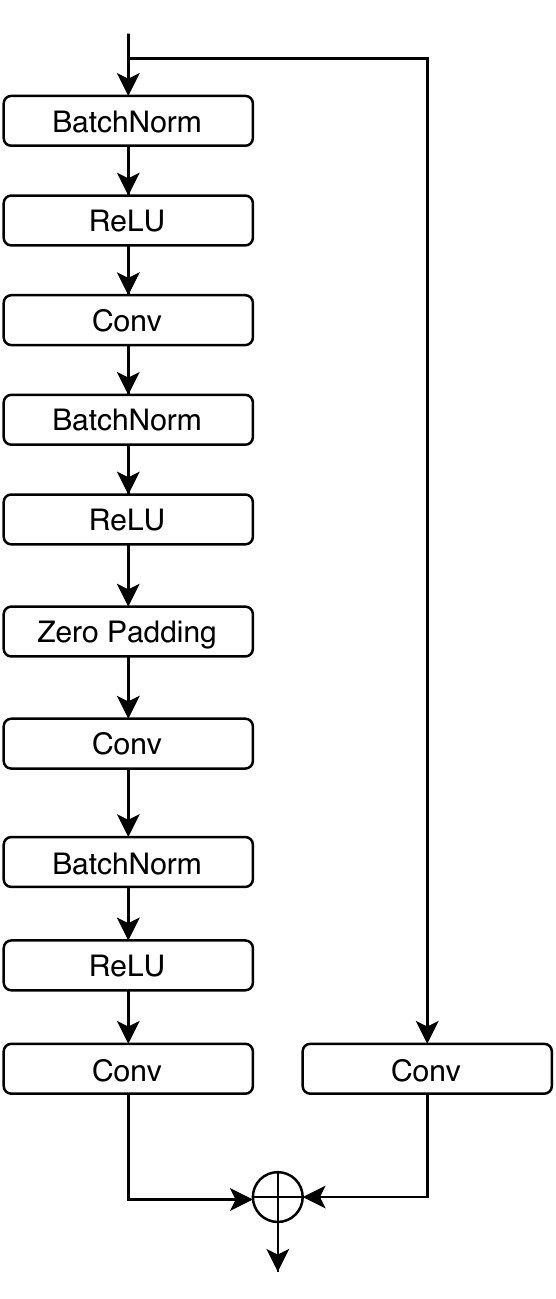}
 \caption{ResNet-E2 is unit 2 of the encoder part.}
 \label{fig:ch4_block_details_b}
 \end{subfigure}
 
 ~ 
 
 \begin{subfigure}[t]{0.49\textwidth}
 \centering
 \includegraphics[height=3.5in]{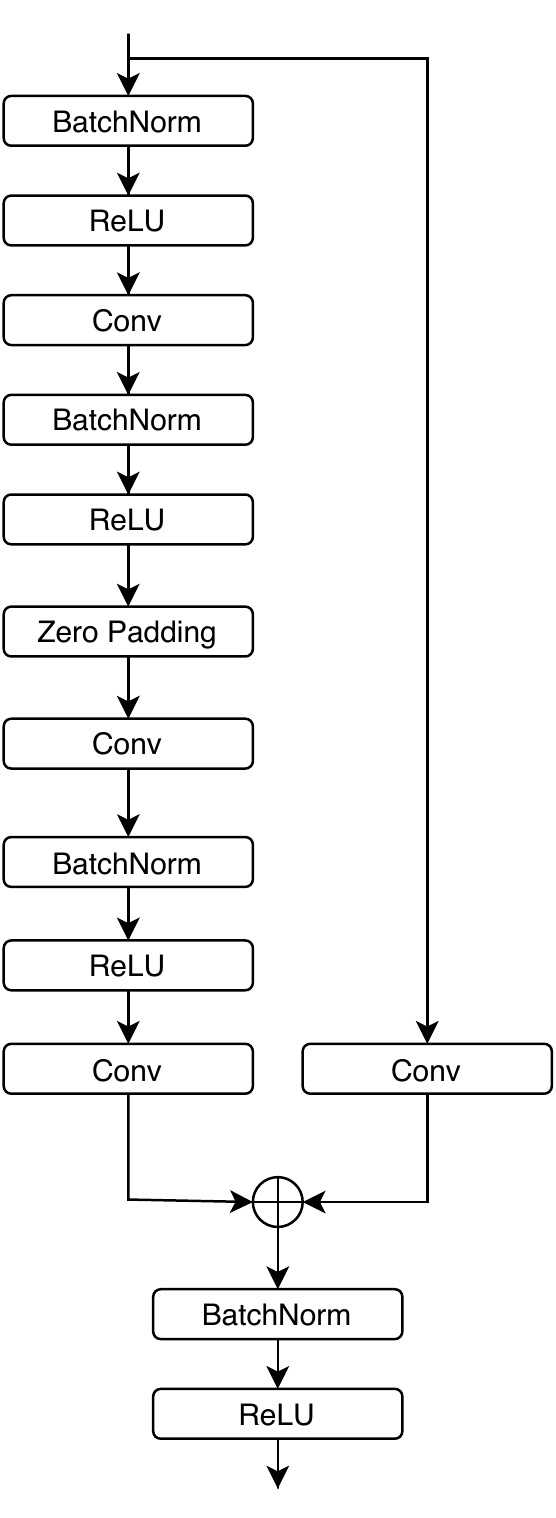}
 \caption{ResNet-E3 is unit 3 of the encoder part.}
 \label{fig:ch4_block_details_c}
 \end{subfigure}%
 ~ 
 \begin{subfigure}[t]{0.49\textwidth}
 \centering
 \includegraphics[height=2in]{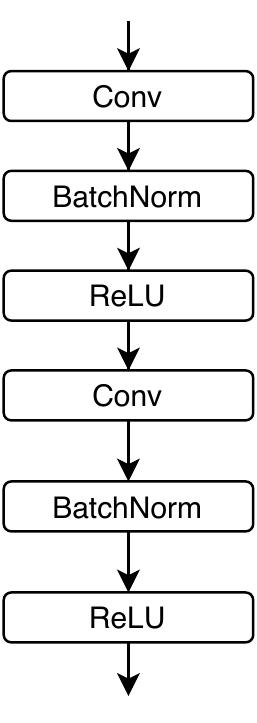}
 \caption{ResNet-D is the block of the decoder part.}
 \label{fig:ch4_block_details_d}
 \end{subfigure}
 
 \caption{Building Blocks of ResNet34-UNet network.}
 \label{fig:block_details}
\end{figure}

Our ResNet34-UNet segmentation network has an encoder and a decoder. The encoder is built by removing the fully connected layer and the global average pooling layer from the end of ResNet-34. The decoder has five blocks, each consisting of a $2 \times 2$ up-sampling layer followed by two sets of layers, each set containing a convolution, batch normalization layer, and Rectified Linear Unit (ReLU) activation layer. In the first four blocks of the decoder, the feature maps after up-sampling are concatenated with the feature maps from the same sized part in the encoder. Finally, we apply a $3 \times 3$ convolution layer followed by sigmoid activation to output the binary masks. The detailed structure of the ResNet34-UNet is shown in Table~\ref{tab:resnet34_unet_arch}. Also, Figure~\ref{fig:block_details} details the building blocks of ResNet34-UNet network.

\begin{table}[t!]
\caption{Architectures of the proposed ResNet34-UNet for pneumothorax segmentation. The symbol [,] denotes the concatenate operation.}
\label{tab:resnet34_unet_arch}
\centering
\begin{tabular}{llll}
\toprule
\textbf{Layer Name} & \textbf{Output Size} & \textbf{Filter Number} & \textbf{Remark} \\
\midrule

Input & $512 \times 512$ & & \\ \midrule

Conv1 & $256 \times 256$ & 64 & kernel Size ($7 \times 7$)\\ 
MaxPool & $128 \times 128$ & & kernel Size ($3 \times 3$) \\ \midrule

\multirow{3}{*}{Encoder1} & \multirow{3}{*}{$128 \times 128$} & \multirow{3}{*}{64} & ResNet-E1 (Figure~\ref{fig:ch4_block_details_a})\\
& & & ResNet-E2 (Figure~\ref{fig:ch4_block_details_b})\\ 
& & & ResNet-E3 (Figure~\ref{fig:ch4_block_details_c})\\ \midrule

\multirow{3}{*}{Encoder2} & \multirow{3}{*}{$64 \times 64$} & \multirow{3}{*}{128} & ResNet-E1 (Figure~\ref{fig:ch4_block_details_a})\\
& & & ResNet-E2 ($x2$) (Figure~\ref{fig:ch4_block_details_b}) \\ 
& & & ResNet-E3 (Figure~\ref{fig:ch4_block_details_c}) \\ \midrule

\multirow{3}{*}{Encoder3} & \multirow{3}{*}{$32 \times 32$} & \multirow{3}{*}{256} & ResNet-E1 (Figure~\ref{fig:ch4_block_details_a})\\
& & & ResNet-E2 ($x4$) (Figure~\ref{fig:ch4_block_details_b}) \\ 
& & & ResNet-E3 (Figure~\ref{fig:ch4_block_details_c}) \\ \midrule

\multirow{3}{*}{Encoder4} & \multirow{3}{*}{$16 \times 16$} & \multirow{3}{*}{512} & ResNet-E1 (Figure~\ref{fig:ch4_block_details_a})\\
& & & ResNet-E2 (Figure~\ref{fig:ch4_block_details_b}) \\ 
& & & ResNet-E3 (Figure~\ref{fig:ch4_block_details_c}) \\ \midrule

\multirow{3}{*}{Decoder1} & \multirow{3}{*}{$32 \times 32$} & & Up-sample1\\
& & 768 & [Up-sample1, Encoder3] \\ 
& & 256 & ResNet-D (Figure~\ref{fig:ch4_block_details_d}) \\ \midrule

\multirow{3}{*}{Decoder2} & \multirow{3}{*}{$64 \times 64$} & & Up-sample2\\
& & 384 & [Up-sample2, Encoder2] \\ 
& & 128 & ResNet-D (Figure~\ref{fig:ch4_block_details_d}) \\ \midrule

\multirow{3}{*}{Decoder3} & \multirow{3}{*}{$128 \times 128$} & & Up-sample3\\
& & 192 & [Up-sample3, Encoder1] \\ 
& & 64 & ResNet-D (Figure~\ref{fig:ch4_block_details_d}) \\ \midrule

\multirow{3}{*}{Decoder4} & \multirow{3}{*}{$256 \times 256$} & & Up-sample4\\
& & 128 & [Up-sample4, Conv1] \\ 
& & 32 & ResNet-D (Figure~\ref{fig:ch4_block_details_d}) \\ \midrule

\multirow{2}{*}{Decoder5} & \multirow{2}{*}{$512 \times 512$} & & Up-sample5\\ 
& & 16 & ResNet-D (Figure~\ref{fig:ch4_block_details_d}) \\ \midrule

Conv2 & $512 \times 512$ & 1 & kernel Size ($3 \times 3$) \\ 

\bottomrule\\
\end{tabular}
\end{table}

\subsection{Training and Testing Details}
We train our network for 100 epochs with batch size 64 on lower resolution ($256 \times 256$) images. Then, the model weights are loaded and the learning rate is initialized to train the model for 70 epochs with batch size 16 on the $512 \times 512$ images. We use the Adam~\cite{adam} optimizer with an initial learning rate of 0.001, which is relatively dropped per epoch using the cosine annealing learning rate technique~\cite{cosinelr}. Also, we apply Stochastic Weight Averaging (SWA)~\cite{swa} to converge more quickly to the wider optima that enables the model to generalize well. The SWA setup for the $256 \times 256$ images has been given for the last five epochs, while for the $512 \times 512$ images, it is for the last three epochs. As a loss function, we use a combination of Binary Cross-Entropy (BCE) and Dice loss. BCE is defined as follows.
\begin{equation}
\text{BCE} = -1/N \sum_{i=1}^{N}[y_{i}\text{log}(p_{i}) + (1 - y_{i})\text{log}(1-p_{i})]
\label{eqn:bce}
\end{equation}
where $N$ is the number of training samples, $y$ is the ground truth value, and $p$ is the predicted value. The Dice loss is defined as follows.
\begin{equation}
\text{DSCL} = 1 - \frac{2 \times |\text{X}\bigcap \text{Y}|}{|\text{X}| + |\text{Y}|}
\label{eqn:dscloss}
\end{equation}
where $X$ is the predicted set of pixels and $Y$ is the ground truth. The expression for the loss function is obtained as follows.
\begin{equation}
\text{BCE-Dice}\; \text{Loss} = \text{BCE} + \text{DSCL}
\label{eqn:total_loss}
\end{equation}

For testing, the images have been resized to $512 \times 512$, and converted to RGB colors with normalization. As for the post-processing steps, we apply horizontal flip, TTA, to test images. Eventually, the final prediction is the average prediction for all images. Moreover, we use the Removal Threshold (R-TH) for each predicted mask to reduce false positives. The components that are less than the minimum size of pixels are removed.

\section{Experiments}
\label{sec:exp}

\subsection{Experiments details}
We conduct several experiments to evaluate our model as shown in Table~\ref{tab:set_exp}. These experiments are tuned based on the validation data. We take from each experiment the best one to evaluate the model on the round-1 test set. Finally, the best experiment is used to infer the round-2 test set.

\begin{table}
\caption{Summary of the settings for all experiments.}
\label{tab:set_exp}
\centering
\begin{tabular}{cccccccc}
\toprule
\multirow{2}{*}[-2pt]{\textbf{Experiment}} & \multirow{2}{*}[-2pt]{\textbf{SWA}} & \multirow{2}{*}[-2pt]{\textbf{B-TH}} & \multirow{2}{*}[-2pt]{\textbf{Channels}} & \multicolumn{2}{c}{\bm{$256 \times 256$}} & \multicolumn{2}{c}{\bm{$512 \times 512$}} \\ \cmidrule{5-8} & & & & \textbf{Batch size} & \textbf{Epochs} & \textbf{Batch size} & \textbf{Epochs} \\ \midrule
Exp1 & \xmark & 0.5 & 1 & 40 & 50 & - & - \\
Exp2 & \cmark & 0.55 & 3 & 40 & 35 & - & - \\ 
Exp3 & \cmark & 0.55 & 3 & 40 & 35 & 14 & 10 \\ 
Exp4 & \cmark & 0.75 & 3 & 40 & 60 & 14 & 29 \\ 
Exp5 & \cmark & 0.75 & 3 & 64 & 100 & 16 & 70 \\ 

\bottomrule \\

\end{tabular}
\end{table}

For the first experiment (Exp1), we train the network at a resolution of $256 \times 256$ and 1-channel for 50 epochs with batch size 40. To enable us to train the network of grayscale images with pre-trained weights, we add the convolution layer to map 1-channel data to 3-channels. We use the Adam optimizer and ReduceLROnPlateau\footnote{\url{https://keras.io/callbacks/\#reducelronplateau}}
scheduler to reduce the learning rate by a factor of 0.2 if no improvement is seen for 5 patience based on the validation loss. The Binarization Threshold (B-TH) is 0.5 for the prediction.

As for the second experiment (Exp2), we train the segmentation network on the $256 \times 256$ resolution images with 3-channels for 35 epochs with 40 batch size. Moreover, the B-TH is 0.55 for the prediction.

In the third experiment (Exp3), we load the weights from the Exp2 model and reinitialize the learning rate. Then, we train the network on the $512 \times 512$ resolution images with 3-channels for 10 epochs with batch size 14. The B-TH is 0.55 for the prediction.

For the fourth experiment (Exp4), we first train our network on the $256 \times 256$ resolution images, and 3-channels for 60 epochs with batch size 40. Then, we load weights and reinitialize the learning rate to retrain the network on the $512 \times 512$ resolution images and 3-channels for 29 epochs with batch size 14. The B-TH is 0.75 for the prediction.

Finally, the configuration of the fifth experiment (Exp5) is the same as Exp4's, except that we train the model on the $256 \times 256$ resolution images for 100 epochs, and $512 \times 512$ for 70 epochs.

We divide the data into 80\% for training and 20\% for validation in the first experiment and 90\% for training and 10\% for validation in the other experiments. All experiments have been trained using the U-Net model with the ResNet-34 backbone. We use SWA with cosine annealing learning rate techniques for all experiments except the first one. As for the prediction, R-TH is 2048, which means that we remove images with a total mask size smaller than 2048 pixels.

Due to the importance and sensitivity of medical data, we have performed a study on the effect of different augmentation techniques on the pneumothorax dataset. For this study, we employ the ResNet34-UNet segmentation model on $256 \times 256$ image resolution and 3-channels. We train the network for 20 epochs and a batch size of 64 using Adam optimizer. For the learning rate, we apply cosine annealing scheduler starting from 1e-4 and gradually decreasing to 1e-5. Also, the value of B-TH is 0.5 and the value of R-TH is 2048.

\subsection{Environment Settings}
All experiments code is written in Python~3. The first four experiments and the experiments of the augmentation study are run on the Colaboratory cloud service provided by Google,\footnote{\url{https://research.google.com/colaboratory/faq.html}} whereas the last experiment is run on a local server with 32 GB of RAM and Nvidia Titan Xp GPU. For Exp5 (2-Stage Training), each training epoch takes about 4 minutes in stage-1 and 8 minutes in stage-2 with a number of trainable parameters of 24.4 million.

\subsection{Metric}
During training, we have used Intersection over Union (IoU)~\cite{jaccard} as a metric. IoU is the area of overlap between the ground truth $(P_{\text{true}})$ and the predicted segmentation $(P_{\text{predicted}})$ divided by the area of union between them. IoU is computed as follows.
\begin{equation}
\text{IoU}(P_{\text{true}}, P_{\text{predicted}}) = \frac{P_{\text{true}} \bigcap P_{\text{predicted}}}{P_{\text{true}} \bigcup P_{\text{predicted}}}
\end{equation}

The competition's official evaluation metric is the mean DSC. This metric is used to compare the pixel-wise agreement between a predicted segmentation and its corresponding ground truth. DSC is computed as follows.
\begin{equation}
\text{DSC}(X, Y) = \frac{2 \times |X\bigcap Y|}{|X| + |Y|}
\end{equation}
where $X$ is the predicted set of pixels and $Y$ is the ground truth.

\section{Results and Discussion}
\label{sec:res}

During round-1 of the competition, the score results reflected the entire round-1 test set, whereas, during round-2 of the competition, the score was calculated with only 1\% of the round-2 test set. Upon the completion of the competition, scores on the private leaderboard were calculated with the remaining 99\% of round-2 test data. 

Table~\ref{tab:val_res} shows the results of the IoU score for all experiments on the validation set with and without using the TTA technique. The results indicate that the use of TTA is highly effective and leads to better predictive performance. We choose the best experiment from each experiment on the validation set to predict the round-1 test set. Table~\ref{tab:st1_res} shows the results of the DSC score for round-1 test set. The 2-Stage Training model achieves a DSC score of 0.8502, and this indicates that 2-Stage Training improves the learning process of the model as it learns the features of its training at different resolutions.

\begin{table}
\caption{IoU score for Round-1 validation set.}
\label{tab:val_res}
\centering
\begin{tabular}{ccc}
\toprule
\multirow{2}{*}[-2pt]{\textbf{Experiment}} & \multicolumn{2}{c}{\textbf{IoU Score}} \\ \cmidrule{2-3} & \textbf{Without TTA} & \textbf{Using TTA} \\ \midrule
 
Exp1 & 0.7242 & 0.7395 \\ 

Exp2 & 0.7349 & 0.7523	\\ 

Exp3 & 0.7105 & 0.7401 \\ 

Exp4 & 0.7422 & 0.7631	\\ 

Exp5 & 0.7638 & 0.7822	\\ 

\bottomrule \\
\end{tabular}
\end{table}

\begin{table}
\caption{DSC score for Round-1 test set.}
\label{tab:st1_res}
\centering
\begin{tabular}{cc}
\toprule
\textbf{Experiment} & \textbf{DSC Score} \\ 
\midrule
 
Exp1 & 0.8125 \\ 

Exp2 &	0.8154 	\\ 

Exp3 & 0.8300 \\ 

Exp4 &	0.8407 	\\ 

Exp5 &	\textbf{0.8502} \\  

\bottomrule \\
\end{tabular}
\end{table}

\begin{figure}
\centering
\includegraphics[width=8cm]{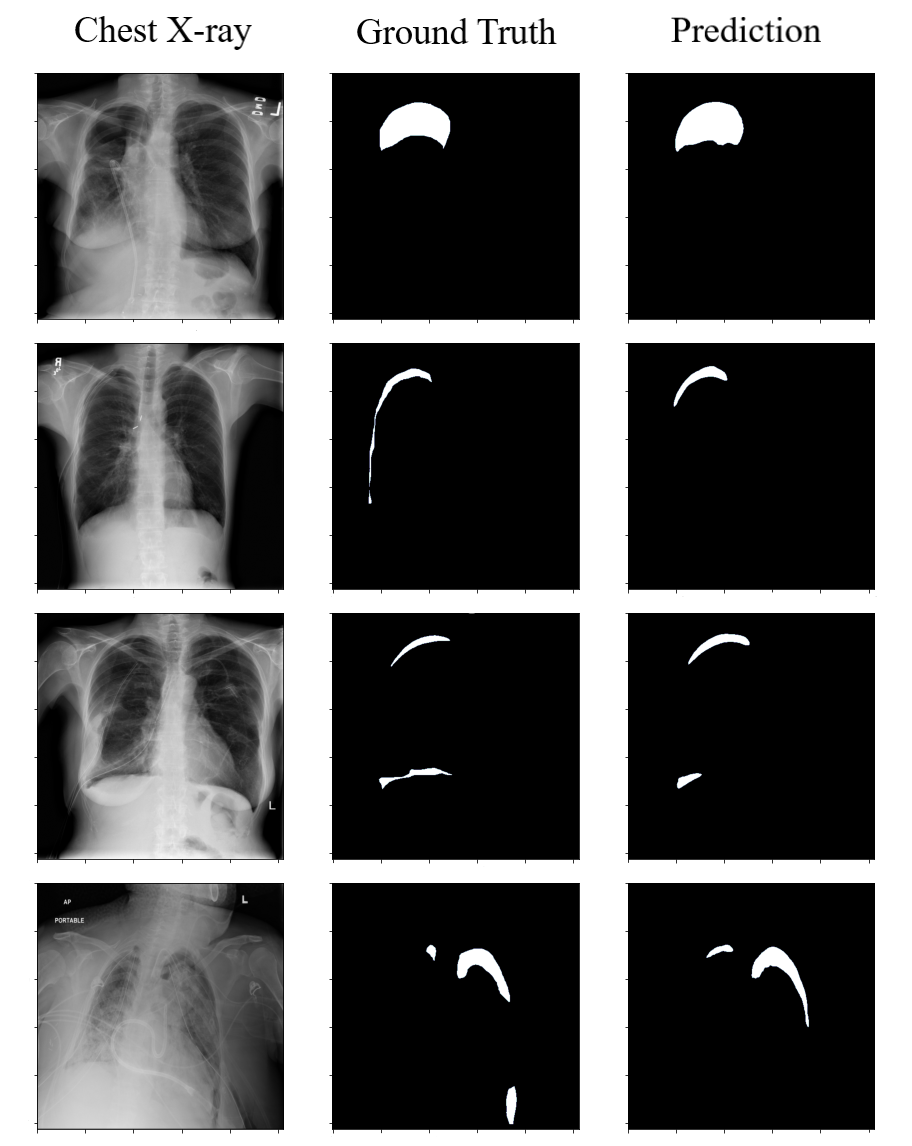}
\caption{Visual segmentation results of the 2ST-UNet on the Round-2 validation set.}
\label{fig:vis_seg_res}
\end{figure}

For round-2 of the competition, a new test was released and the training set was updated to include a round-1 test set. We choose the best experiment (Exp5) 2-Stage Training to infer the round-2 test set. This experiment achieves the highest score on both the validation set and the round-1 test set using the TTA technique. Table~\ref{tab:val_rse2} shows the IoU score for the 2ST-UNet on round-2 validation set with TTA and R-TH configurations. We can see that the use of TTA and R-TH improves the performance of the model for the segmentation task. Figure~\ref{fig:vis_seg_res} shows the visual segmentation results of the 2ST-UNet on round-2 validation set.

\begin{table}
\caption{IoU score for the 2ST-UNet of Round-2 validation set.}
\label{tab:val_rse2}
\centering
\begin{tabular}{ccc}
\toprule

\textbf{TTA} & \textbf{R-TH} & \textbf{IoU Score}\\ \midrule

\xmark & \xmark & 0.7543 \\
\xmark & \cmark & 0.7693\\ 
\cmark & \xmark & 0.7945\\ 
\cmark & \cmark & 0.7946 \\
\bottomrule\\
\end{tabular}
\end{table}

As shown in Table~\ref{tab:st2_res}, we infer the round-2 test set using the model that has been trained on the round-1 training set. Then, we train the model on the round-2 training set to predict the round-2 test set. We also make a comparison using TTA and without using it to confirm the effectiveness of our approach. The results indicate that the use of the TTA technique improves the prediction, where the horizontal flip transformation of the test images increases the chances of capturing the target shape and predicting performance. Also, we calculate the average prediction time per sample for 100 times, and 2ST-UNet takes 25.6 milliseconds and this speed in diagnosing pneumothorax can be helpful in the medical fields to start treatment. Our model achieves a DSC score of 0.8356. 

\begin{table}
\caption{Round-2 results of 2ST-UNet system with DSC score.}
\label{tab:st2_res}
\centering
\begin{tabular}{cccc}
\toprule
\multirow{2}{*}{\textbf{Train set}} & \multirow{2}{*}{\textbf{TTA}} & \multicolumn{2}{c}{\textbf{DSC Score}} \\ \cline{3-4} & & \textbf{Public Leaderboard} & \textbf{Private Leaderboard} \\
\midrule
 
Round-1 Train & \xmark & 0.9005 & 0.8247 \\ 

Round-1 Train 	& \cmark &	0.8938	& 0.8330	\\ 

Round-2 Train & \xmark & 0.9011 & 0.8298 \\ 

Round-2 Train & \cmark &\textbf{0.9023}	& \textbf{0.8356}	\\ 

\bottomrule \\
\end{tabular}
\end{table}

Table~\ref{tab:aug} shows the groups of augmentation techniques that we study and apply. As shown in Table~\ref{tab:study_aug}, each of the A1, A3, and A4 augmentation groups improve the segmentation performance of the validation set and the test set compared to the Non-Augmentation case (no data augmentation is performed). The A2 augmentation group is worse than the Non-Augmentation case, while the combination of the A2 and A1 achieves a higher score on the validation set and the test set compared to both the Non-Augmentation case and A1. Without TTA and R-TH, A4 has achieved the highest IoU score of 0.7660 on the validation set, while A4 has achieved the highest DSC score of 0.8012 on the test set. Using TTA, A4 has achieved the highest IoU score of 0.7656 on the validation set, while the combination of A1 and A2 has achieved the highest DSC score of 0.8092 on the test set. Using R-TH, A4 has achieved the highest IoU score of 0.7660 on the validation set, while A3 has achieved the highest DSC score of 0.8172 on the test set. Using TTA and R-TH together, A1 has achieved the highest IoU score of 0.7850 on the validation set, while the combination of A1 and A2 has achieved the highest DSC score of 0.8212 on the test set. Based on these results, we conclude that all of the augmentation groups that have used in this study boost the performance of pneumothorax segmentation by helping the model to learn the correct features and patterns.

\begin{table}
\caption{Augmentation Techniques.}
\label{tab:aug}
\centering
\begin{tabular}{cc}
\toprule
\textbf{Augmentation Group} & \textbf{Augmentation Technique} \\ 
\midrule
 
A1 & Horizontal Flip \\ 

A2 & Random Contrast, Random Gamma, and Random Brightness \\ 

A3 & Elastic Transform, Optical Distortion, Grid Distortion \\ 

A4 & Random Sized Crop 	\\ 

\bottomrule \\
\end{tabular}
\end{table}

\begin{table}
\caption{Experimental results of the different data augmentation techniques in terms of IoU for the round-2 validation set and DSC for the round-2 test set.}
\label{tab:study_aug}
\centering
\begin{tabular}{ccccc}
\toprule

\textbf{Augmentation Group} & \textbf{TTA} & \textbf{R-TH} & \textbf{Round-2 Validation IoU} & \textbf{Round-2 Test DSC}\\ 
\midrule
 
\multirow{4}{*}{Non-Augmentation} & \xmark & \xmark & 0.7459 & 0.7859 \\
& \cmark & \xmark & 0.7370 & 0.7824\\ 
& \xmark & \cmark & 0.7840 & 0.8111 \\ 
& \cmark & \cmark & 0.7832 & 0.8074\\ 
\midrule

\multirow{4}{*}{A1} & \xmark & \xmark & 0.7490 & 0.8000\\
& \cmark & \xmark & 0.7593 & 0.8067\\ 
& \xmark & \cmark & 0.7842 & 0.8127 \\ 
& \cmark & \cmark & 0.7850 & 0.8188\\ 
\midrule

\multirow{4}{*}{A2} & \xmark & \xmark & 0.7330 & 0.7820 \\
& \cmark & \xmark & 0.7341 & 0.7767\\ 
& \xmark & \cmark & 0.7825 & 0.8105 \\ 
& \cmark & \cmark & 0.7819 & 0.8105\\ 
\midrule

\multirow{4}{*}{A1 $+$ A2} & \xmark & \xmark & 0.7496 & 0.8004 \\
& \cmark & \xmark & 0.7608 & 0.8092\\ 
& \xmark & \cmark & 0.7842 & 0.8152 \\ 
& \cmark & \cmark & 0.7843 & 0.8212\\ 
\midrule

\multirow{4}{*}{A3} & \xmark & \xmark & 0.7457 & 0.7985 \\
& \cmark & \xmark & 0.7379 & 0.7889\\ 
& \xmark & \cmark & 0.7846 & 0.8172 \\ 
& \cmark & \cmark & 0.7833 & 0.8095 \\ 
\midrule

\multirow{4}{*}{A4} & \xmark & \xmark & 0.7660 & 0.8012 \\
& \cmark & \xmark & 0.7656 & 0.8027\\ 
& \xmark & \cmark & 0.7844 & 0.8142 \\ 
& \cmark & \cmark & 0.7842 & 0.8144\\ 

\bottomrule \\
\end{tabular}
\end{table}

Table~\ref{tab:com_res} shows the performance of our method compared to the winning methods of the Pneumothorax Challenge on the test dataset. We notice that, for this task, the high resolution images increases the performance in addition to the use of deeper and more powerful backbones. It is worth mentioning that we have been awarded a bronze medal on Kaggle for being in the top 9\% of competitors with a rank of 124 out of 1,475.

\begin{table}
\caption{DSC score compared with the top winning methods on the pneumothorax Round-2 test set.}
\centering
\label{tab:com_res}
\begin{tabular}{cccc}
\toprule
\textbf{Team} & \textbf{Model} & \textbf{Image size} & \textbf{DSC Score} \\ 
\midrule
 
[dsmlkz] sneddy & U-Net & $1024 \times 1024$ & 0.8679 \\
    
X5 & Deeplabv3+, U-Net & $1024 \times 1024$ & 0.8665 \\ 

2ST-UNet & U-Net & $512 \times 512$ & 0.8356\\ 

\bottomrule \\
\end{tabular}
\end{table}

\section{Conclusion}
\label{sec:conc}

In this paper, a 2-Stage Training system is proposed to segment pneumothorax based on an elegant state-of-the-art architecture called U-Net. The backbone used is ResNet-34 pre-trained on the ImageNet dataset and the dataset consists of chest X-ray images provided by the 2019 SIIM-ACR Pneumothorax Segmentation Challenge. We have employed data augmentation, SWA and TTA techniques to improve the network's predictions. Our method achieves a mean DSC of 0.8356 and was ranked 124 out of 1,475 competitors.

\section*{Acknowledgement}
We gratefully acknowledge the support of the Deanship of Research at the Jordan University of Science and Technology for supporting this work via Grant \#20190180 in addition to NVIDIA Corporation for the donation of the Titan Xp GPU used for this research.

\bibliographystyle{unsrt}  
\bibliography{references} 

\end{document}